\documentclass[journal]{IEEEtran}

\usepackage[bookmarks,colorlinks]{hyperref}
\usepackage{enumitem}
\usepackage[linesnumbered,ruled,lined]{algorithm2e}

\usepackage{booktabs,array,tabularx,makecell,threeparttable}
\newcolumntype{C}[1]{>{\centering\arraybackslash}m{#1}}
\newcolumntype{Y}{>{\centering\arraybackslash}X}

\AtBeginDocument{}

\usepackage{graphicx}
\usepackage{dcolumn}
\usepackage{bm}

\usepackage[usenames,dvipsnames]{xcolor}

\usepackage{amsmath,amsfonts}
\usepackage{mathtools} 

\usepackage{array}
\usepackage[caption=false,font=normalsize,labelfont=sf,textfont=sf]{subfig}
\usepackage{textcomp}
\usepackage{stfloats}
\usepackage{url}
\usepackage{verbatim}
\usepackage{graphicx}
\usepackage{cite}
\usepackage{svg}
\usepackage[noend]{algpseudocode} 
\usepackage{etoolbox}
\usepackage{booktabs}   
\usepackage{array}      
\usepackage{multirow}

\usepackage{float}
\usepackage{pgfplots}
\usepackage[bookmarks,colorlinks]{hyperref}
\usepackage[linesnumbered,ruled,lined]{algorithm2e}
\usepackage{enumitem}
\usepackage{algpseudocode}
\usetikzlibrary{shapes.multipart,intersections}
\usepackage{cite}
\usepackage{amsmath,amssymb,amsfonts,amsthm,steinmetz}
\usepackage{graphicx}
\usepackage{mathrsfs}  
\usepackage{textcomp}
\usepackage{acronym}
\usepackage{xcolor}
\usepackage{upgreek,xspace}

\usepackage{tikz}
\usetikzlibrary{calc}
\makeatletter
\newcommand{\gettikzxy}[3]{%
  \tikz@scan@one@point\pgfutil@firstofone#1\relax
  \edef#2{\the\pgf@x}%
  \edef#3{\the\pgf@y}%
}
\makeatother

\usepackage[draft]{todonotes}

\usepackage{esvect}

\usepackage{times}
\usepackage{bm}
\usepackage{amsmath}
\usepackage{amssymb}
\usepackage{stmaryrd}
\usepackage{babel}
\usepackage{graphics, graphicx}
\usepackage{xcolor}
\usepackage{gensymb}
\usepackage{cite}
\usepackage{enumitem}
\usepackage{url}

\ifCLASSINFOpdf

\else

\fi

\usepackage[all=normal,paragraphs=normal,floats=normal,mathspacing=normal,wordspacing=normal,charwidths=tight,mathdisplays=normal,leading=normal]{savetrees}

\begin{document}

\title{Effective Electromagnetic Degrees of Freedom \\in Backscatter MIMO Systems}

\author{Philipp~del~Hougne,~\IEEEmembership{Member,~IEEE}
\thanks{This work was supported in part by the Nokia Foundation (project 20260028), the ANR France 2030 program (project ANR-22-PEFT-0005), the ANR PRCI program (project ANR-22-CE93-0010), the Rennes M\'etropole AES program (project ``SRI''), the European Union's European Regional Development Fund, and the French Region of Brittany and Rennes M\'etropole through the contrats de plan \'Etat-R\'egion program (projects ``SOPHIE/STIC \& Ondes'' and ``CyMoCoD'').}
\thanks{P.~del~Hougne is with Aalto University, Department of Electronics and Nanoengineering, 02150 Espoo, Finland and Univ Rennes, CNRS, IETR - UMR 6164, F-35000, Rennes, France (e-mail: philipp.del-hougne@univ-rennes.fr).}
}

\maketitle

\begin{abstract}
While the definition of the effective electromagnetic degrees of freedom (EEMDOFs) of a static linear multiple-input multiple-output (MIMO) system is well established, the counterpart for a backscatter MIMO (BS-MIMO) system is so far missing. A BS-MIMO system encodes the input information into the loads of backscatter elements. Due to mutual coupling, the mapping from load configuration to observed fields is fundamentally non-linear, which complicates the analysis of BS-EEMDOFs. We introduce a definition of BS-EEMDOFs based on the Jacobian of the observed fields with respect to the load configuration. We derive a closed-form expression from multiport network theory which demonstrates that the number of BS-EEMDOFs is fundamentally a distributed variable, whose distribution depends on the mutual coupling between the backscatter elements and the coherent illumination. The modes associated with BS-EEMDOFs lie in the column space of the end-to-end channel matrix from backscatter array ports to receiver ports, but the number of BS-EEMDOFs is generally different from the number of benchmark EEMDOFs associated with the same array being coherently fed rather than tunably terminated. The dependence on the coherent illumination yields optimized coherent illumination as a control knob for the number of BS-EEMDOFs.
We present numerical and experimental results for the evaluation and optimization of the number of BS-EEMDOFs in different radio environments with reconfigurable intelligent surfaces.
\end{abstract}


\IEEEpeerreviewmaketitle

\section{Introduction}
\label{sec_introduction}

Effective degrees of freedom quantify how many independent directions in the space of input parameters produce distinguishable changes of the outputs. The effective electromagnetic degrees of freedom (EEMDOFs) of a static linear multiple-input multiple-output (MIMO) system play an important role in electromagnetic information theory~\cite{franceschetti2009capacity,miller2019waves,di2024electromagnetic}. The number of EEMDOFs is commonly defined as the number of significant singular values of the matrix characterizing the MIMO system (i.e., the \textit{linear mapping} from input wavefronts to output wavefronts)~\cite{shiu2000fading,miller2000communicating,piestun2000electromagnetic,poon2005degrees,migliore2006role,muharemovic2008antenna,yuan2021electromagnetic,pizzo2022landau,yuan2023effects,ruiz2023degrees}. 
This approach assumes hence that input information is encoded into the input wavefront, and that output information is extracted from the output wavefront. However, it is also conceivable that input information is encoded into the loads of an array of backscatter elements, while the output information is again extracted from the output wavefront. We refer to the latter as a backscatter MIMO (BS-MIMO) system; to be clear, our distinction between MIMO and BS-MIMO resides in how the input information is encoded. To the best of our knowledge, so far there exists no rigorous technique for evaluating the backscatter EEMDOFs (BS-EEMDOFs) of a BS-MIMO system. The crux lies in the \textit{non-linear mapping} from load configuration to output wavefront in BS-MIMO, due to mutual coupling (MC) between the backscatter elements. 

The principle of backscatter communications is known at least since the 1940s~\cite{stockman1948communication,brooker2013lev}, and nowadays  omnipresent in everyday life in the form of RFID technology. One variant known as ``ambient'' backscatter communications leverages pre-existing ambient signals to communicate~\cite{liu2013ambient,van2018ambient}. The recent emergence of large arrays of backscatter elements, referred to as programmable metasurfaces or reconfigurable intelligent surfaces (RISs), enables massive multi-element backscatter communications that encode the input information into the backscatter elements' load configuration~\cite{cui2019direct,basar2019media,tang2020mimo,zhao2020metasurface,li2021single}. Moreover, wave-domain physical neural networks can deliberately exploit the ``structural non-linearity'' of the input-to-output mapping in massive backscatter systems~\cite{momeni2023backpropagation}. 

Some recent works have considered the EEMDOFs of MIMO systems involving RISs, either treating the RIS as a transmitter capable of generating arbitrary aperture fields in the spirit of holographic MIMO~\cite{dardari_JSAC,decarli2021communication}, or treating the RIS as a programmable scattering object to maximize the EEMDOFs of an end-to-end wireless MIMO channel~\cite{del2019optimally,del2019optimized,do2022line,ruizWSA}; however, an analysis treating the RIS as an array of backscatter elements whose load configurations encode the input information is missing. Consequently, to the best of our knowledge, the literature contains no definition or technique for evaluating BS-EEMDOFs.

To tackle the definition of backscatter EEMDOFs (BS-EEMDOFs), we take inspiration from other domains confronted with quantifying the diversity of non-linear input-to-output mappings. For compressed sensing with non-linear observations,~\cite{blumensath2013compressed} proposed to approximate the non-linear mapping with an affine Taylor-series-type approximation in the vicinity of a reference point. Related ideas were also explored to analyze deep neural networks~\cite{wang2016analysis,mardani2019degrees}. The approach of~\cite{blumensath2013compressed} was recently leveraged in~\cite{del2025low} to quantify the diversity of a sensing operator in wireless multiport sensing. Specifically,~\cite{del2025low} considered the multiplexing of the information carried by the scattering matrix of a load network terminating an array of backscatter elements across different configurations of a programmable over-the-air fixture onto an observed single-input single-output wireless end-to-end channel. However, the focus in~\cite{del2025low} was on assessing identifiability in an inverse problem, rather than analyzing information transmission in BS-MIMO from the point of view of electromagnetic information theory, as we do in the present work. Moreover,~\cite{del2025low} only considered one small-scale backscatter system and focused on configurational diversity rather than spatial diversity. 

In addition, beyond the definition and evaluation of BS-EEMDOFs, the following two intriguing questions were not raised or tackled prior to the present work. \textit{First}, how are the BS-EEMDOFs of a given antenna array terminated by arbitrarily tunable loads related to the EEMDOFs of the same antenna array when it is coherently excited via its feeds? Are BS-EEMDOFs bounded by EEMDOFs or can they exceed EEMDOFs? \textit{Second}, what role does the coherent illumination of a backscatter array play for the BS-EEMDOFs? Can one optimize the backscatter array's illumination to maximize the BS-EEMDOFs?
While existing literature has considered optimizing the coherent illumination of a backscatter array for various end-to-end metrics~\cite{zawawi2018multiuser,mishra2019multi,mishra2019sum,al2020massive}, these works neither account for MC between backscatter elements nor do they examine BS-EEMDOFs and whether they can be maximized with an optimized coherent illumination. 

In this paper, we fill the research gaps on BS-EEMDOFs and answer these questions. Our contributions are summarized as follows:
\begin{enumerate}
    \item We generalize the definition of EEMDOFs as the distribution of the effective number of singular values (participation number) of the Jacobian of the output wavefront with respect to the parameters encoding the input information. 
    \item We derive a closed-form expression of the relevant Jacobian based on a multiport-network model for BS-MIMO to define BS-EEMDOFs.
    \item We compare operating a given antenna array in a given radio environment as a conventional transmit array (coherently excited via its feeds) or as a backscatter array (feeds terminated by tunable loads). We show that the column space of the Jacobian for the backscatter scenario lies in the column space of the fixed end-to-end channel matrix in the conventional scenario; however, the singular value distributions are generally different; for the backscatter case, it depends on the load configuration and the backscatter array illumination. Thus, in contrast to conventional EEMDOFs, BS-EEMDOFs are generally characterized by a distribution rather than a single value.
    \item We optimize the fixed coherent wavefront illuminating the backscatter array to minimize or maximize the mean of the distribution of BS-EEMDOFs. 
    \item Based on a full-wave simulation, we evaluate and optimize the distribution of BS-EEMDOFs for a 64-element RIS parametrizing a $3\times4$ MIMO channel in a wireless network-on-chip, and compare it to the number of conventional EEMDOFs for the same array. We further examine the influence of the choice of the distributions of the loads on the distribution of BS-EEMDOFs. 
    \item Based on experimental measurements, we  evaluate and optimize the distribution of BS-EEMDOFs for a 100-element RIS parametrizing a $4\times 4$ MIMO channel in four distinct radio environments.
\end{enumerate}

\textit{Organization:} 
In Sec.~\ref{sec_theory}, \textit{first}, we generalize the definition of effective degrees of freedom, \textit{second}, we apply it to a conventional MIMO system as a sanity check, and, \textit{third}, we apply it to a BS-MIMO system using multiport network theory (MNT) and compare it to a conventional MIMO system.
In Sec.~\ref{sec_numerics}, we leverage a full-wave numerical simulation of a wireless network-on-chip to evaluate and optimize BS-EEMDOFs, to examine the influence of the distribution of loads, and to evaluate the number of conventional EEMDOFs.
In Sec.~\ref{sec_experiments}, we leverage experimental measurements to evaluate and optimize BS-EEMDOFs in four distinct radio environments.
In Sec.~\ref{sec_conclusion}, we close with a brief conclusion and outlook.

\textit{Notation:} 
$\mathbf{A} = \mathrm{diag}(\mathbf{a})$ denotes the diagonal matrix $\mathbf{A}$ whose diagonal entries are $\mathbf{a}$.
$\mathbf{I}_a$ denotes the $a \times a$ identity matrix.
$\left[ \mathbf{A}^{-1} \right]_\mathcal{BC}$ denotes the block of $\mathbf{A}^{-1}$ selected by the sets of indices $\mathcal{B}$ and $\mathcal{C}$. 
$\mathbf{A}^\dagger$ denotes the conjugate transpose of $\mathbf{A}$.

\section{Theory}
\label{sec_theory}

\subsection{Generalized definition of effective degrees of freedom}
\label{subsec_genEDOF}

Effective degrees of freedom quantify how many independent directions in the space of the $N_\mathrm{in}$ input parameters $\boldsymbol{\theta}_\mathrm{in}\in\mathbb{C}^{N_\mathrm{in}}$ produce distinguishable changes of the $N_\mathrm{out}$ outputs $\boldsymbol{\theta}_\mathrm{out}\in\mathbb{C}^{N_\mathrm{out}}$. The input-to-output mapping is characterized by a function $\boldsymbol{\theta}_\mathrm{out} = f(\boldsymbol{\theta}_\mathrm{in})$ that may be non-linear. Nonetheless, locally around a reference point $\boldsymbol{\theta}_{\mathrm{in},0}$, an affine approximation of $f$ can be formulated based on its Jacobian~\cite{blumensath2013compressed}:
\begin{equation}
  \boldsymbol{\theta}_\mathrm{out}
  \;\approx\;
  \boldsymbol{\theta}_{\mathrm{out},0}
  +
  \mathbf{J}(\boldsymbol{\theta}_{\mathrm{in},0})
  \,\delta\boldsymbol{\theta}_\mathrm{in},
\end{equation}
where $\boldsymbol{\theta}_{\mathrm{out},0} = f(\boldsymbol{\theta}_{\mathrm{in},0})$,  $\delta\boldsymbol{\theta}_\mathrm{in} = \boldsymbol{\theta}_\mathrm{in} - \boldsymbol{\theta}_{\mathrm{in},0}$, and
\begin{equation}
  \mathbf{J}(\boldsymbol{\theta}_{\mathrm{in},0})   =
  \left.
  \frac{\partial \boldsymbol{\theta}_\mathrm{out}}{\partial \boldsymbol{\theta}_\mathrm{in}}
  \right|_{\boldsymbol{\theta}_\mathrm{in}=\boldsymbol{\theta}_{\mathrm{in},0}} \in \mathbb{C}^{N_\mathrm{out}\times N_\mathrm{in}}.
\end{equation}
Around the reference point $\boldsymbol{\theta}_{\mathrm{in},0}$, the effective degrees of freedom are given by the effective number of singular values of $\mathbf{J}(\boldsymbol{\theta}_{\mathrm{in},0})$. Based on a singular value decomposition $\mathbf{J}(\boldsymbol{\theta}_{\mathrm{in},0})   =\mathbf{U}\,\boldsymbol{\Sigma}\,\mathbf{V}^\dagger$, where $\mathbf{U}\in\mathbb{C}^{N_\mathrm{out}\times {\tilde{N}}}$, $\mathbf{V}\in\mathbb{C}^{N_\mathrm{in}\times {\tilde{N}}}$, $\boldsymbol{\Sigma}=\mathrm{diag}(\sigma_1,\dots,\sigma_{\tilde{N}})$, $\tilde{N}=\mathrm{min}(N_\mathrm{in},N_\mathrm{out})$, and $\sigma_1\geq\dots\geq\sigma_{\tilde{N}}\geq 0$, we evaluate the Jacobian's participation number:
\begin{equation}
    M(\boldsymbol{\theta}_{\mathrm{in},0}) =   \frac{\bigl(\sum_{i=1}^{\tilde{N}} \sigma_i^2\bigr)^2}{\sum_{i=1}^{\tilde{N}} \sigma_i^4}.
    \label{eq3}
\end{equation}
By construction, $1 \leq M(\boldsymbol{\theta}_{\mathrm{in},0}) \leq {\tilde{N}}$; $M(\boldsymbol{\theta}_{\mathrm{in},0})$ attains its upper limit of ${\tilde{N}}$ if all singular values are non-zero and equal.
Throughout this paper, we mean the participation number defined in~(\ref{eq3}) when referring to the number of effective degrees of freedom.

Because $M(\boldsymbol{\theta}_{\mathrm{in},0})$ generally depends on the reference point $\boldsymbol{\theta}_{\mathrm{in},0}$, the effective degrees of freedom are characterized by the distribution of $M(\boldsymbol{\theta}_{\mathrm{in},0})$ over a set of operating points of interest. 
$M(\boldsymbol{\theta}_{\mathrm{in},0})$ is thus a random variable whose distribution depends on the distribution of the operating points of interest $\boldsymbol{\theta}_{\mathrm{in},0}$ across the relevant domain of input parameters.

\begin{figure}
    \centering
    \includegraphics[width=\columnwidth]{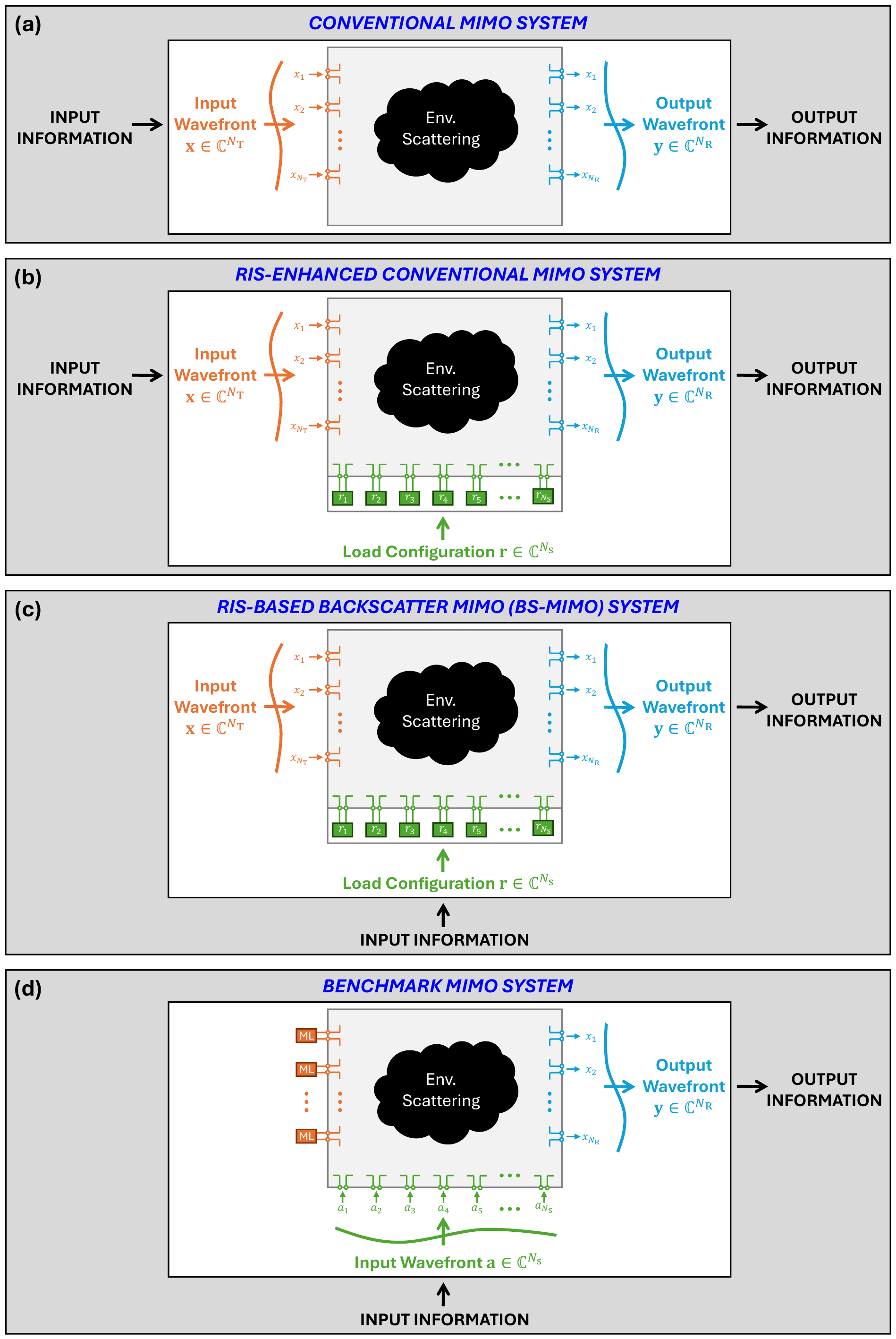}
    \caption{Taxonomy of considered MIMO systems. (a) In a \textit{conventional MIMO system}, the input information is encoded into the input wavefront $\mathbf{x}$, and the output information is extracted from the output wavefront $\mathbf{y}$. (b) In an \textit{RIS-enhanced conventional MIMO system}, the configuration $\mathbf{r}$ of an RIS is optimized to shape the end-to-end MIMO transfer function. (c) In an \textit{RIS-based backscatter MIMO (BS-MIMO) system}, the input information is encoded into the RIS configuration $\mathbf{r}$ (rather than into the input wavefront $\mathbf{x}$), while the output information is again extracted from the output wavefront $\mathbf{y}$. (d) In a \textit{benchmark MIMO system} for the BS-MIMO system from (c) we remove the RIS loads and inject a coherent wavefront $\mathbf{a}$ carrying the input information via the RIS array, while  extracting the output information again from the output wavefront $\mathbf{y}$. ML denotes matched load. }
    \label{Fig1}
\end{figure}

\subsection{Sanity check: Application to a conventional MIMO system}

For the conventional MIMO system depicted in Fig.~\ref{Fig1}(a) with $N_\mathrm{T}$ transmitters, $N_\mathrm{R}$ receivers, and end-to-end channel matrix $\mathbf{H}$, we identify:
\begin{subequations}
\begin{equation}
  N_\mathrm{in} \leftrightarrow N_\mathrm{T}; \  N_\mathrm{out} \leftrightarrow N_\mathrm{R},
\end{equation}
\begin{equation}
    \boldsymbol{\theta}_{\mathrm{in}} \leftrightarrow \mathbf{x}  ; \   \boldsymbol{\theta}_{\mathrm{out}}  \leftrightarrow \mathbf{y},
\end{equation}
\begin{equation}
    \mathbf{y}=f(\mathbf{x})=\mathbf{H}\mathbf{x}.
\end{equation}
\end{subequations}
Evaluating the Jacobian, we obtain
\begin{equation}
  \mathbf{J}(\mathbf{x}_{0})   =
  \left.
  \frac{\partial \mathbf{y}}{\partial \mathbf{x}}
  \right|_{\mathbf{x}=\mathbf{x}_{0}} = \mathbf{H} \in \mathbb{C}^{N_\mathrm{R}\times N_\mathrm{T}},
\end{equation}
where we note the absence of any dependence on $\mathbf{x}$. For this reason, the distribution of $M(\mathbf{x}_0)$ has in fact no dependence on $\mathbf{x}_0$ and collapses to a single value. Consequently, the effective degrees of freedom of a conventional MIMO system are simply the participation number of the end-to-end wireless channel matrix, which is the well-established result.

\subsection{Application to a BS-MIMO system}

We now apply the generalized definition of effective degrees of freedom from Sec.~\ref{subsec_genEDOF} to a BS-MIMO system. To that end, we define our system model in Sec.~\ref{subsubsec_system_model} and then proceed with the evaluation of the BS-EEMDOFs in Sec.~\ref{subsubsec_evaluation}. Subsequently, we closely examine the result from Sec.~\ref{subsubsec_evaluation} in Sec.~\ref{subsubsec_examinationBSEEMDOFs}.

\subsubsection{System Model}
\label{subsubsec_system_model}

We consider the scenario depicted in Fig.~\ref{Fig1}(b,c) comprising $N_\mathrm{T}$ transmitting antennas, $N_\mathrm{R}$ receiving antennas, and $N_\mathrm{S}$ backscatter elements. Each backscatter element is an antenna whose port is terminated by a tunable load. The scattering between these $N=N_\mathrm{T}+N_\mathrm{R}+N_\mathrm{S}$ ports is assumed to be static, linear, and passive; it is characterized by the scattering matrix $\mathbf{S}\in\mathbb{C}^{N\times N}$ that is defined using the same reference impedance $Z_0$ at each port. $\mathbf{S}$ captures the environmental scattering within the radio environment as well as the structural scattering of the antennas. We assume that all signal generators and detectors are matched to $Z_0$.
The only tunable components in this setup are the loads of the backscatter elements. The load of the $i$th element has a reflection coefficient $r_i$, and we collect all $N_\mathrm{S}$ reflection coefficients in $\mathbf{r}=[r_1,r_2,\dots,r_{N_\mathrm{S}}]\in\mathbb{C}^{N_\mathrm{S}}$.

According to MNT~\cite{anderson_cascade_1966,ha1981solid,prod2024efficient}, the dependence of the end-to-end channel matrix $\mathbf{H}\in\mathbb{C}^{N_\mathrm{R}\times N_\mathrm{T}}$ on $\mathbf{S}$ and $\mathbf{r}$ is given by
\begin{equation}
    \mathbf{H}(\mathbf{r}) =  \mathbf{S}_\mathcal{RT}
    + \mathbf{S}_\mathcal{RS}
      \left( \mathbf{I}_{N_\mathrm{S}} - \mathbf{\Phi}(\mathbf{r}) \mathbf{S}_\mathcal{SS} \right)^{-1} \mathbf{\Phi}(\mathbf{r})
      \mathbf{S}_\mathcal{ST},
    \label{eq1}
\end{equation}
where $\mathbf{\Phi}(\mathbf{r}) = \mathrm{diag}(\mathbf{r})$ and $\mathcal{R}$, $\mathcal{S}$ and $\mathcal{T}$ denote, respectively, the sets of port indices associated with receiving antennas, backscatter elements, and transmitting antennas. The non-linear dependence of $\mathbf{H}$ on $\mathbf{r}$ is apparent.

The output wavefront $\mathbf{y}\in\mathbb{C}^{N_\mathrm{R}}$ at the ports of the receiving antennas is
\begin{equation}
    \mathbf{y}(\mathbf{r},\mathbf{x})=\mathbf{H}(\mathbf{r})\mathbf{x},
    \label{eq2}
\end{equation}
where $\mathbf{x}\in\mathbb{C}^{N_\mathrm{T}}$ is the input wavefront injected into the ports of the transmitting antennas. We emphasize that $\mathbf{y}$ depends linearly on $\mathbf{x}$ but non-linearly on $\mathbf{r}$. 
In an RIS-enhanced conventional MIMO systems as depicted in Fig.~\ref{Fig1}(b), the input information is encoded into $\mathbf{x}$ and the output information is extracted from $\mathbf{y}$, implying a \textit{linear} input-to-output mapping characterized by $\mathbf{H}$; by choosing the loads of the backscatter elements, $\mathbf{H}$ can be optimized. 
In contrast, in an RIS-based BS-MIMO as depicted in Fig.~\ref{Fig1}(c), the input information is encoded into $\mathbf{r}$ and the output information is extracted from $\mathbf{y}$, implying a \textit{non-linear} input-to-output mapping; by choosing the input wavefront $\mathbf{x}$, one can optimize the non-linear mapping (see below).

For the purpose of our BS-EEMDOF analysis, we assume that the loads terminating the backscatter elements are mutually independent and continuously tunable, so that each reflection coefficient $r_i$ can take any complex value within a fixed disk centered at the origin. This idealization removes quantization constraints on the loads and mirrors conventional EEMDOF analysis, which similarly neglects RF-chain quantization and focuses on the properties of the end-to-end wireless MIMO channel rather than on specific RF-chain hardware. In the same spirit, we analyze our BS-MIMO system while remaining agnostic to the detailed implementation of the load modulators. This idealization is technically plausible, as impedance modulators capable of fine-grained, independent control of phase and amplitude have been demonstrated in practice~\cite{correia2017quadrature,belo2019iq,zhu2025millimeter}.

\subsubsection{Evaluation of BS-EEMDOFs}
\label{subsubsec_evaluation}
For the BS-MIMO system described in Sec.~\ref{subsubsec_system_model} and displayed in Fig.~\ref{Fig1}(c), we identify:
\begin{subequations}
\begin{equation}
  N_\mathrm{in} \leftrightarrow N_\mathrm{S}; \  N_\mathrm{out} \leftrightarrow N_\mathrm{R},
\end{equation}
\begin{equation}
     \boldsymbol{\theta}_{\mathrm{in}}  \leftrightarrow \mathbf{r} ; \    \boldsymbol{\theta}_{\mathrm{out}} \leftrightarrow \mathbf{y},
\end{equation}
\begin{align}
    \mathbf{y}&=f(\mathbf{r},\mathbf{x})\\&=\left[\mathbf{S}_\mathcal{RT}
    + \mathbf{S}_\mathcal{RS} \left(  \mathbf{I}_{N_\mathrm{S}} - \mathbf{\Phi}(\mathbf{r})\mathbf{S}_\mathcal{SS} \right)^{-1}\mathbf{\Phi}(\mathbf{r})  \mathbf{S}_\mathcal{ST}\right] \mathbf{x}.
\end{align}
\end{subequations}

Evaluating the Jacobian (see Appendix for details), we obtain
\begin{equation}
  \begin{aligned}
  \mathbf{J}(\mathbf{r}_0,\mathbf{x})
   &=
  \left.
  \frac{\partial \mathbf{y}(\mathbf{r},\mathbf{x})}
       {\partial \mathbf{r}}
  \right|_{\mathbf{r}=\mathbf{r}_0} \\
   &=
  \mathbf{S}_\mathcal{RS}\,
  \mathbf{G}(\mathbf{r}_0)\,
  \operatorname{diag}\bigl( \mathbf{W}(\mathbf{r}_0) \,\mathbf{x} \bigr)
  \;\\&\in\;
  \mathbb{C}^{N_\mathrm{R}\times N_\mathrm{S}},
  \end{aligned}
  \label{eq:JBS_full}
\end{equation}
where
\begin{subequations}
\begin{equation}
   \mathbf{W}(\mathbf{r}) =  \mathbf{S}_\mathcal{SS}\,\mathbf{G}(\mathbf{r})\,
              \boldsymbol{\Phi}(\mathbf{r})\,
              \mathbf{S}_\mathcal{ST}
              + \mathbf{S}_\mathcal{ST},
              \label{eq_def_W}
\end{equation}
\begin{equation}
    \mathbf{G}(\mathbf{r})   =   \bigl(\mathbf{I}_{N_\mathrm{S}} - \boldsymbol{\Phi}(\mathbf{r})\,  \mathbf{S}_\mathcal{SS}\bigr)^{-1}.
\end{equation}
\end{subequations}
We note that the Jacobian depends on both $\mathbf{r}_0$ and $\mathbf{x}$. The BS-EEMDOFs are thus characterized by the distribution of the participation number of $\mathbf{J}(\mathbf{r},\mathbf{x})$ across the relevant domains of $\mathbf{r}$ and $\mathbf{x}$.

\subsubsection{Examination of BS-EEMDOFs} 
\label{subsubsec_examinationBSEEMDOFs}
We now derive a few important insights from inspecting~(\ref{eq:JBS_full}). 

\textit{First}, as depicted in Fig.~\ref{Fig1}(d), let us imagine that the backscatter elements are excited by a controllable coherent signal via their feeds instead of being terminated by tunable loads. Then, compared to Fig.~\ref{Fig1}(c), the same antenna array in the same radio environment would act as an $N_\mathrm{S}$-element transmit array in a conventional $N_\mathrm{R}\times N_\mathrm{S}$ MIMO scheme whose end-to-end channel matrix is $\mathbf{S}_\mathcal{RS}$.\footnote{We do not need to worry about the antennas with port indices in $\mathcal{T}$ because they are excited by a fixed signal that does not carry any information.} 
The number of EEMDOFs of this conventional MIMO scheme is thus given by the participation number of $\mathbf{S}_\mathcal{RS}$ and cannot be altered without changing the array and/or the radio environment. 
The modes associated with the EEMDOFs lie in the column space of $\mathbf{S}_\mathcal{RS}$.\footnote{The column space of a matrix $\mathbf{A} \in \mathbb{C}^{m\times n}$ is the subspace spanned by its columns, i.e., $\operatorname{col}(\mathbf{A}) = \{\mathbf{A}\mathbf{x} : \mathbf{x} \in \mathbb{C}^n\}
= \operatorname{span}\{\mathbf{a}_1,\dots,\mathbf{a}_n\}$, where $\mathbf{a}_i$ denotes
the $i$th column of $\mathbf{A}$.}

\textit{Second}, we return to the BS-MIMO scheme in Fig.~\ref{Fig1}(c). Defining $\mathbf{B}(\mathbf{r}_0,\mathbf{x})=\mathbf{G}(\mathbf{r}_0)\,
    \operatorname{diag}\bigl(\mathbf{W}(\mathbf{r}_0)\,\mathbf{x}\bigr)\in\mathbb{C}^{N_\mathrm{S}\times N_\mathrm{S}}$, we find
\begin{equation}
    \mathbf{J}(\mathbf{r}_0,\mathbf{x}) = \mathbf{S}_\mathcal{RS}\,\mathbf{B}(\mathbf{r}_0,\mathbf{x}).
    \label{eq_colspace}
\end{equation}
By inspection of (\ref{eq_colspace}), we observe that the column space of $\mathbf{J}(\mathbf{r}_0,\mathbf{x})$ must lie within the column space of $\mathbf{S}_\mathcal{RS}$.
In other words, the modes associated with BS-EEMDOFs lie in the same space as the modes associated with the conventional EEMDOFs. However, the non-zero singular values of $\mathbf{J}(\mathbf{r}_0,\mathbf{x})$ are those of the product $\mathbf{S}_\mathcal{RS}\,\mathbf{B}(\mathbf{r}_0,\mathbf{x})$, which generically differ from the singular values of $\mathbf{S}_\mathcal{RS}$ itself. As a consequence, the number of BS-EEMDOFs (i.e., the participation number of $\mathbf{J}(\mathbf{r}_0,\mathbf{x})$) is in general different from the number of conventional EEMDOFs (i.e., the participation number of $\mathbf{S}_\mathcal{RS}$) for the same physical array in the same radio environment.

\textit{Third}, because the number of BS-EEMDOFs depends on the coherent illumination $\mathbf{x}$ of the backscatter elements, optimizing $\mathbf{x}$ opens up a new control knob. By optimizing $\mathbf{x}$ (under an appropriate power constraint), we can adjust the relative weights of the columns of $\mathbf{J}(\mathbf{r}_0,\mathbf{x})$ at a given operating point $\mathbf{r}_0$ and thereby optimize the singular value distribution of the Jacobian. Ultimately, we can thus optimize the number of BS-EEMDOFs. 
More generally, we can optimize a fixed $\mathbf{x}$ (i.e., independent of the operating point $\mathbf{r}$) to shape desired characteristics (e.g., the mean or median) of the distribution of the number of BS-EEMDOFs. 

\textit{Fourth}, in the absence of MC between the backscatter elements (i.e., with $\mathbf{S}_\mathcal{SS}=\mathbf{0}$),~(\ref{eq:JBS_full}) simplifies to 
\begin{equation}
    \mathbf{J}^\mathrm{noMC}(\mathbf{x}) = \mathbf{S}_\mathcal{RS} \, \mathrm{diag}(\mathbf{S}_\mathcal{ST}\mathbf{x}),
\end{equation}
which is independent of $\mathbf{r}_0$ but still depends on $\mathbf{x}$. Thus, the distribution of the number of BS-EEMDOFs collapses to a single value only in the absence of MC and with a fixed coherent illumination.

To summarize, we have established the following points:
\begin{enumerate}
    \item Both the modes associated with BS-EEMDOFs and the modes associated with EEMDOFs lie in the column space  of the end-to-end channel matrix from the backscatter array ports to the receiver ports.
    \item The number of BS-EEMDOFs generally differs from the number of EEMDOFs.
    \item The number of BS-EEMDOFs generally depends on $\mathbf{r}$ (unless MC between backscatter elements is negligible) and $\mathbf{x}$ (unless $\mathbf{x}$ is fixed).
    \item The number of BS-EEMDOFs is generally a distributed random variable unless MC between backscatter elements is negligible \textit{and} $\mathbf{x}$ is fixed.
    \item The properties of the distribution of the number of BS-EEMDOFs can be optimized by controlling $\mathbf{x}$. 
\end{enumerate}

\section{Full-Wave Numerical Results}
\label{sec_numerics}

In this section, we apply the theoretical insights from Sec.~\ref{sec_theory} to a numerically simulated BS-MIMO system. A single full-wave numerical simulation can provide all MNT model parameters (free of any ambiguity) by replacing lumped tunable elements with lumped ports~\cite{tapie2023systematic,Naffouri2024,almunif2025network}. 
In contrast, as discussed in Sec.~\ref{sec_experiments}, in most BS-MIMO experiments the MNT model parameters can only be obtained indirectly and with inevitable ambiguities, which precludes a direct application of the theoretical analysis from Sec.~\ref{sec_theory}. Thus, the full-wave numerical analysis in this section goes beyond what can be analyzed in Sec.~\ref{sec_experiments} based on experimental measurements.

\begin{figure}
    \centering
    \includegraphics[width=\columnwidth]{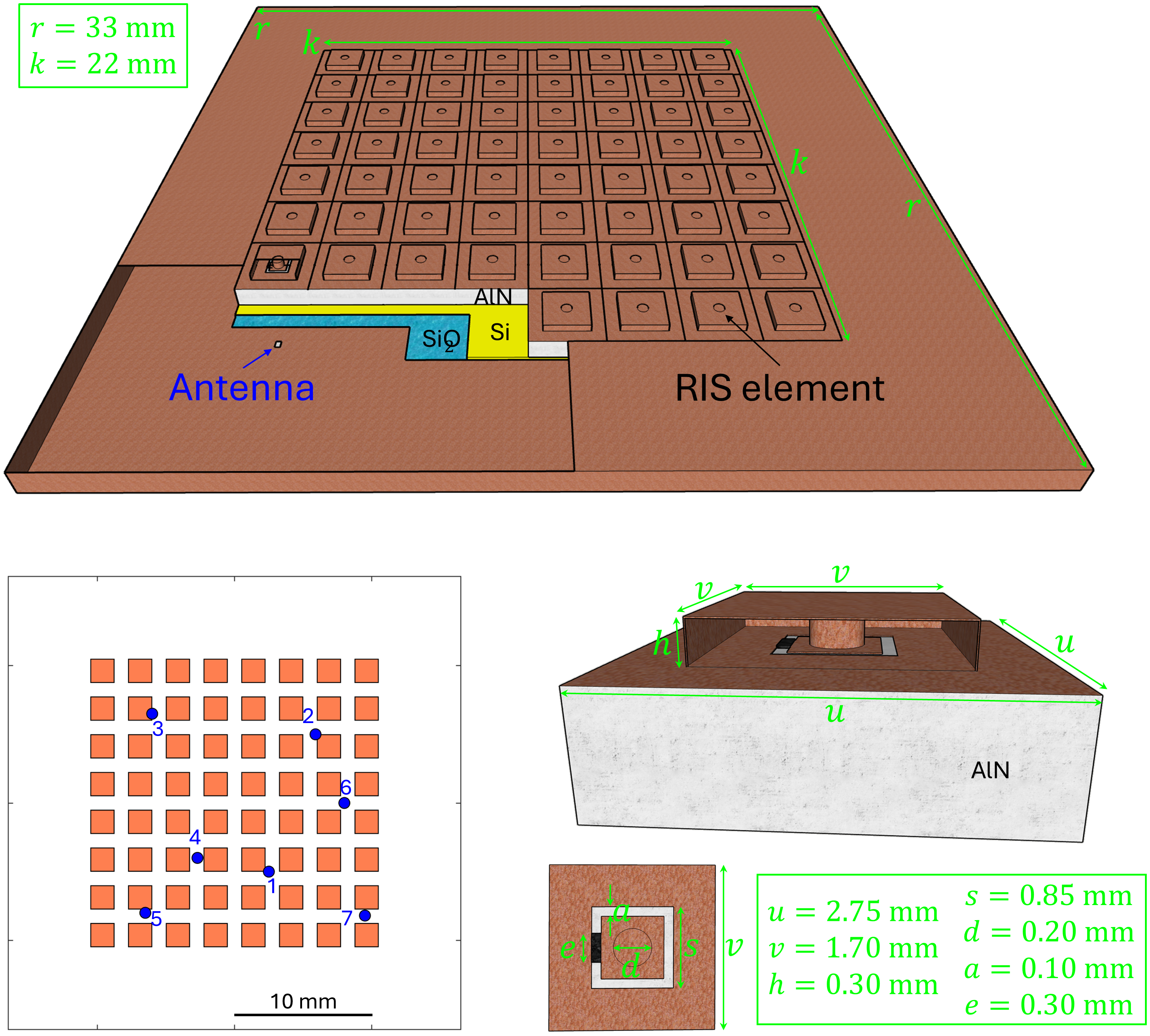}
    \caption{Numerical setup from~\cite{tapie2023systematic} used in Sec.~\ref{sec_numerics}. The setup comprises 7 antennas and 64 RIS elements.}
    \label{Fig2}
\end{figure}

Our numerical setup is the RIS-parametrized wireless network-on-chip (WNoC) described in~\cite{tapie2023systematic}, comprising 64 RIS elements operating in the 60~GHz regime. The only difference with respect to~\cite{tapie2023systematic} is that we consider 7 instead of only 4 antennas. To be clear, our contribution in this section is the application of the theoretical analysis from Sec.~\ref{sec_theory} to this numerical setup, but not the design and simulation of the setup itself (which were already reported in~\cite{tapie2023systematic}). Thus, we do not dwell on the design and simulation of the setup here; we reproduce relevant aspects in Fig.~\ref{Fig2} and refer interested readers to~\cite{tapie2023systematic} for further details.

Our analysis in this section is for an operating frequency of 60~GHz. Given that the setup includes 7 antennas, we choose $N_\mathrm{T}=3$ and $N_\mathrm{R}=4$. We consider two different choices of how to partition the 7 antennas into receiving and transmitting ones:
\begin{itemize}
    \item $\mathcal{T}=\{2,4,5\}$ and $\mathcal{R}=\{1,3,6,7\}$,
    \item $\mathcal{T}=\{2,6,7\}$ and $\mathcal{R}=\{1,3,4,5\}$.
\end{itemize}
Evaluating the distribution of the number of BS-EEMDOFs requires specifying the choice of the backscatter array's coherent illumination $\mathbf{x}\in\mathbb{C}^{3}$ and the constraints on the distribution of random realizations of the backscatter array's load configuration $\mathbf{r}\in\mathbb{C}^{64}$. Regarding the former, we consider three distinct distributions of $\mathbf{x}$:
\begin{itemize}
    \item \textit{RAND}: For each sample, we draw a new realization of $\mathbf{x}$ uniformly at random from the complex unit sphere.
    \item \textit{MAX}: We use the same $\mathbf{x}_\mathrm{MAX}$ for all samples; the fixed coherent illumination is chosen to \textit{maximize} the mean of the distribution of BS-EEMDOFs. 
    \item \textit{MIN}: We use the same $\mathbf{x}_\mathrm{MIN}$ for all samples; the fixed coherent illumination is chosen to \textit{minimize} the mean of the distribution of BS-EEMDOFs. 
\end{itemize}
Regarding the latter, we consider three distinct constraints under which the entries of $\mathbf{r}$ are randomly drawn:
\begin{itemize}
    \item \textit{PIN}: We assume that the entries of $\mathbf{r}$ are constrained to two possible values corresponding to the two possible reflection coefficients at 60~GHz deduced from the data sheet of a commercial PIN diode. Following the analysis in~\cite{tapie2023systematic}, the two possible values are $r_\mathrm{ON}=-0.8116$ and $r_\mathrm{OFF}=0.6366-0.7712\jmath$.  
    \item \textit{PM}: We assume that the entries of $\mathbf{r}$ are constrained to two idealized possible values: $r_\mathrm{ON}=1$ and $r_\mathrm{OFF}=-1$.  
    \item \textit{UNI}: We assume that the entries of $\mathbf{r}$ are continuously tunable, and the distribution of the magnitudes is uniform on [$0,1$] while the distribution of the phases is uniform on [$0,2\pi$). 
\end{itemize}

To identify $\mathbf{x}_\mathrm{MAX}$ and $\mathbf{x}_\mathrm{MIN}$, we define the metric to be maximized or minimized as the mean of the distribution of the participation number of the Jacobian defined in~(\ref{eq:JBS_full}), where the random variable generating the distribution is the random realization of $\mathbf{r}$. Specifically, we estimate our metric for a given $\mathbf{x}$ based on 1500 random realizations of $\mathbf{r}$ (subject to one of the three aforementioned constraints). For each realization of $\mathbf{r}$, we evaluate the Jacobian in closed form using~(\ref{eq:JBS_full}). Our optimization is carried out with a gradient-free Nelder–Mead algorithm~\cite{lagarias1998convergence} applied to a real-valued parametrization of $\mathbf{x}$. The choice of a gradient-free algorithm is motivated by the lack of a closed-form mapping from $\mathbf{x}$ to the metric to be optimized (i.e., the mean of the distribution of the number of BS-EEMDOFs).
To avoid poor local optima, we use a multistart strategy: the Nelder–Mead search is run from three random initial illuminations uniformly drawn on the complex unit sphere, and we retain the solution that yields the best metric. 

Intuitively, $\mathbf{x}_\mathrm{MAX}$ seeks to balance the Jacobian's singular values (on average) to maximize the number of BS-EEMDOFs, which is a natural goal to pursue from the perspective of information transfer. Meanwhile, $\mathbf{x}_\mathrm{MIN}$ pursues the opposite goal by seeking to make the Jacobian's singular values as unbalanced as possible (on average). Besides satisfying a fundamental curiosity about understanding the limits of tunability of the distribution of the number of BS-EEMDOFs, minimizing the BS-EEMDOFs may be the goal in adversarial attacks aimed at disrupting a BS-MIMO link. Explorations of optimizing other statistical characteristics of the distributions, such as their spread, with optimized coherent illumination are deferred to future work.

\begin{figure}
    \centering
    \includegraphics[width=\columnwidth]{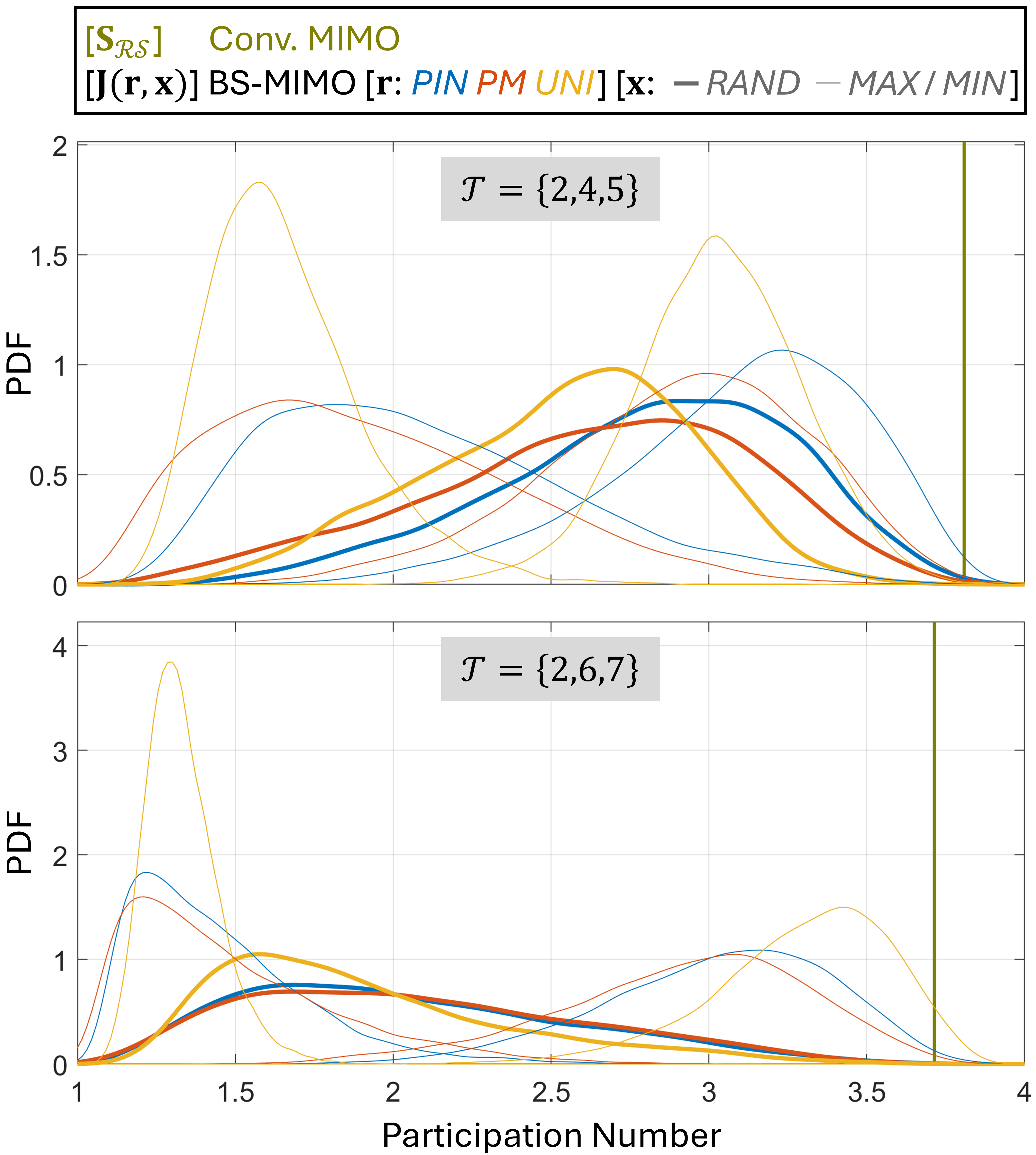}
    \caption{Probability density functions (PDFs) of the number of BS-EEMDOFs in the numerical setup from Fig.~\ref{Fig2} for different constraints on the entries of $\mathbf{r}$ (different line colors), different choices of $\mathbf{x}$ (different line thicknesses; \textit{MAX}~vs.~\textit{MIN} distinction by inspection), and different choices of $\mathcal{T}$ (different panels). In addition, the number of conventional EEMDOFs for the same array in the same radio environment is shown as a vertical line. Each displayed PDF is evaluated based on 10,000 samples using~(\ref{eq:JBS_full}).}
    \label{Fig3}
\end{figure}

Our results are displayed in Fig.~\ref{Fig3}. For the two considered choices of $\mathcal{T}$, the number of conventional EEMDOFs is 3.81 and 3.71, respectively, which is close to the upper bound of $\tilde{N}=4$. This makes sense because of the rich-scattering nature of the radio environment and because $N_\mathrm{S}=64 \gg N_\mathrm{R}=4$. Meanwhile, the number of BS-EEMDOFs is a distributed variable. The averages and standard deviations of the distributions displayed in Fig.~\ref{Fig3} are summarized in Table~\ref{table1}. In all cases, the standard deviations are substantial, emphasizing that the distributed nature of the BS-EEMDOFs is significant. The distributions tend to be narrower with fixed coherent illumination, which makes sense because the random process underlying the generation of the distributions then only depends on the statistics of $\mathbf{r}$. The significant width of the distributions for fixed coherent illumination is a direct consequence of MC between the RIS elements. Indeed, as discussed at the end of Sec.~\ref{subsubsec_examinationBSEEMDOFs}, with fixed coherent illumination the distribution of the number of BS-EEMDOFs on $\mathbf{r}$ would collapse to a single value if MC was negligible. The substantial width of the distributions with fixed coherent illumination in Fig.~\ref{Fig3} thus emphasizes the importance of accounting for MC in evaluating BS-EEMDOFs.
Moreover, we note that none of the averages of the distributions in Fig.~\ref{Fig3} exceeds the number of conventional EEMDOFs. However, this observation is likely specific to the considered setup and should not be taken as a generalized conclusion. In principle, it seems conceivable that the mean of the distribution of BS-EEMDOFs exceeds the number of conventional EEMDOFs, especially with optimized coherent illumination.

\begin{table}[t]
    \centering
    \caption{Mean and standard deviation of the distributions of the \\number of BS-EEMDOFs shown in Fig.~\ref{Fig3}.}
    \label{table1}
    \begin{threeparttable}
    \begin{tabular}{ccccc}
        \toprule
          &   & \textit{RAND} & \textit{MAX} & \textit{MIN} \\
        \midrule
        \multirow{3}{*}{$\{2,4,5\}$}
            & \textit{PIN} & $2.80 \pm 0.46$ & $3.10 \pm 0.40$ & $2.09 \pm 0.48$ \\
            & \textit{PM}  & $2.62 \pm 0.52$ & $2.90 \pm 0.41$ & $1.91 \pm 0.46$ \\
            & \textit{UNI} & $2.53 \pm 0.42$ & $3.03 \pm 0.26$ & $1.66 \pm 0.24$ \\
        \midrule
        \multirow{3}{*}{$\{2,6,7\}$}
            & \textit{PIN} & $2.06 \pm 0.53$ & $3.03 \pm 0.36$ & $1.45 \pm 0.28$ \\
            & \textit{PM}  & $2.08 \pm 0.53$ & $2.91 \pm 0.40$ & $1.48 \pm 0.33$ \\
            & \textit{UNI} & $1.91 \pm 0.47$ & $3.30 \pm 0.28$ & $1.33 \pm 0.12$ \\
        \bottomrule
    \end{tabular}
    \end{threeparttable}
\end{table}

We now take a closer look at the dependence of the distribution on $\mathcal{T}$, $\mathbf{r}$ and $\mathbf{x}$. 
With \textit{RAND} illumination, the average of the distributions is markedly higher for $\mathcal{T}=\{2,4,5\}$ than for $\mathcal{T}=\{2,6,7\}$ in all cases. 
With \textit{MAX} [\textit{MIN}] illumination, however, the averages are comparable or higher [lower] for $\mathcal{T}=\{2,6,7\}$ than for $\mathcal{T}=\{2,4,5\}$. The flexibility to shape the distributions with optimized coherent illumination is thus dependent on $\mathcal{T}$. The largest flexibility is seen for $\mathcal{T}=\{2,6,7\}$ with the \textit{UNI} constraint on the loads, where the average of the distribution can be tuned from 1.33 (with \textit{MIN}) to 3.30 (with \textit{MAX}).
These results demonstrate that by merely optimizing the coherent illumination $\mathbf{x}$, one can nearly recover the conventional EEMDOFs or, conversely, strongly suppress BS-EEMDOFs, without changing the array hardware or environment. This is in sharp contrast to conventional EEMDOFs which are fixed for a given array hardware and environment.
We further observe that the distributions tend to be narrower in the case of the \textit{UNI} constraint on the loads. However, the qualitative dependence of the distributions on $\mathbf{x}$ is the same for all considered constraints on $\mathbf{r}$ and choices of $\mathcal{T}$.

\section{Experimental Results}
\label{sec_experiments}

In this section, we experimentally evaluate and optimize the number of BS-EEMDOFs in four distinct radio environments. All four experiments involve the same 4-element transmit array, 4-element receive array, and 100-element RIS (with 1-bit-programmable elements); the four radio environments differ regarding the strength of environmental scattering, which in turn influences the strength of MC between the RIS elements~\cite{rabault2024tacit}. The major challenge in experimental work on BS-EEMDOFs relates to the lack of direct access to the MNT model parameters. The latter include $\mathbf{S}$ as well as the two possible values of the entries of $\mathbf{r}$.

A direct measurement of the MNT model parameters would require that:
\begin{enumerate}[label=(\roman*)]
    \item The backscatter array has a \textit{modular} design in which its $N_\mathrm{S}$ ports are connectorized.
    \item The total number of ports $N=N_\mathrm{T}+N_\mathrm{R}+N_\mathrm{S}$ does not exceed the number of ports of the available vector network analyzer (VNA).
\end{enumerate}
In our experiments (which are representative of a typical BS-MIMO setup), neither of these conditions is satisfied:
\begin{enumerate}[label=(\roman*)]
    \item Our RIS has an \textit{integrated} design in which the tunable lumped elements (PIN diodes) are integrated into the RIS elements; it is not possible to disconnect the PIN diodes from the RIS elements in order to connect the RIS elements (or the PIN diodes) via coaxial cables to VNA ports.
    \item $N=4+4+100=108$ drastically exceeds the number of ports of our eight-port VNA.
\end{enumerate}
Thus, neither $\mathbf{S}$ nor the two possible values of the entries of $\mathbf{r}$ can be measured directly in our experiment.

Obtaining the MNT model parameters based on a full-wave simulation of our setup is also not feasible.  \textit{First}, the detailed geometry and material composition of our radio environment is unknown and the RIS may deviate from its intended design due to fabrication inaccuracies. \textit{Second}, even if perfect knowledge of the system details was available, the computational cost of simulating this electrically very large system would be prohibitive. 

The only option to proceed consists thus in estimating the MNT model parameters based on backscatter measurements. Recent work on a ``Virtual VNA'' technique demonstrates that an unambiguous estimation of $\mathbf{S}$ is possible provided that the ports of the backscatter array can be terminated by three distinct and known individual loads, as well as known coupled loads~\cite{del2024virtual,del2025virtual,v2na3p0,kitvna}.\footnote{Very recently,~\cite{del2025wireless} noticed that two distinct known individual loads and known coupled loads can be sufficient because the latter provide effectively a third termination distinct from the two individual loads.} However, our RIS elements can only be terminated by two distinct individual loads whose reflection coefficients we assume to be unknown. 

For our RIS composed of 1-bit-programmable elements with unknown states (which are representative characteristics of typical RIS prototypes), we can nonetheless experimentally estimate a set of \textit{proxy} MNT model parameters that accurately predict the end-to-end wireless channel $\mathbf{H}$ as a function of the chosen RIS configuration control vector~\cite{sol2024experimentally,multibitRIS,del2025ambiguity,del2025reducedrank}. The \textit{proxy} MNT model parameters are ambiguous; there is an infinite number of sets of proxy MNT parameters that accurately map a RIS control vector to the corresponding end-to-end channel matrix. A discussion of the ambiguities in the possible values of the entries of $\mathbf{r}$ can be found in~\cite{multibitRIS}, and a discussion of the ambiguities in $\mathbf{S}$ assuming known loads can be found in the Virtual VNA literature~\cite{del2024virtual,del2025virtual,v2na3p0}. In this section, we estimate a set of proxy MNT model parameters for each of our four experimental setups using the technique described in~\cite{multibitRIS}.

While the ambiguities in the MNT model parameters are immaterial for optimizing the RIS configuration with regard to an objective based on the accurately predicted end-to-end wireless channel matrix~\cite{sol2024experimentally,multibitRIS,del2025ambiguity,del2025reducedrank}, the analysis in Sec.~\ref{sec_theory} is based on the model parameters rather than the end-to-end channel matrix. We cannot directly apply the analysis presented in Sec.~\ref{sec_theory} to an experimentally estimated proxy MNT model because different sets of proxy parameters would not yield the same values of EEMDOFs and BS-EEMDOFs. In this section, we thus modify the analysis from Sec.~\ref{sec_theory}. 
Specifically, in the experiments we do not evaluate the closed-form Jacobian in~(\ref{eq:JBS_full}); instead, we approximate the relevant Jacobian numerically by finite differences with respect to the RIS control variables, using only end-to-end channel matrices predicted by the proxy MNT model.\footnote{In the present case with 1-bit-programmable RIS elements, evaluating the finite-difference Jacobian with respect to the loads' reflection coefficients or with respect to the loads' control variables yields Jacobians that differ only by a global nonzero complex scalar prefactor, which does not influence the Jacobian's participation number.} Since the mapping from the RIS control vector to the end-to-end MIMO channel is unambiguous, the resulting BS-EEMDOFs are independent of the particular ambiguous parametrization of the MNT model.
Moreover, we limit the considered constraints on $\mathbf{r}$ to the one that matches our hardware (i.e., \textit{PIN} using the two values of the possible entries of $\mathbf{r}$ that we estimate jointly with the other MNT proxy parameters~\cite{multibitRIS}). Furthermore, we do not evaluate the conventional EEMDOFs of our RIS in the four radio environments because we do not unambiguously know $\mathbf{S}_\mathcal{RS}$.

Despite their ambiguities, our experimentally estimated proxy MNT models are still highly valuable for the work reported in this section. Without a high-accuracy proxy model, every iteration of the optimization to find $\mathbf{x}_\mathrm{MAX}$ or $\mathbf{x}_\mathrm{MIN}$ would require $\mathcal{O}(N_\mathrm{S})$ experimental measurements. Meanwhile, the corresponding measurements can be accurately predicted by our proxy model in a computationally highly efficient manner leveraging the Woodbury identity~\cite{prod2023efficient}. Thus, experimentally estimating an ambiguous proxy MNT model for each experimentally considered setup is very valuable.

\begin{figure*}
    \centering
    \includegraphics[width=2\columnwidth]{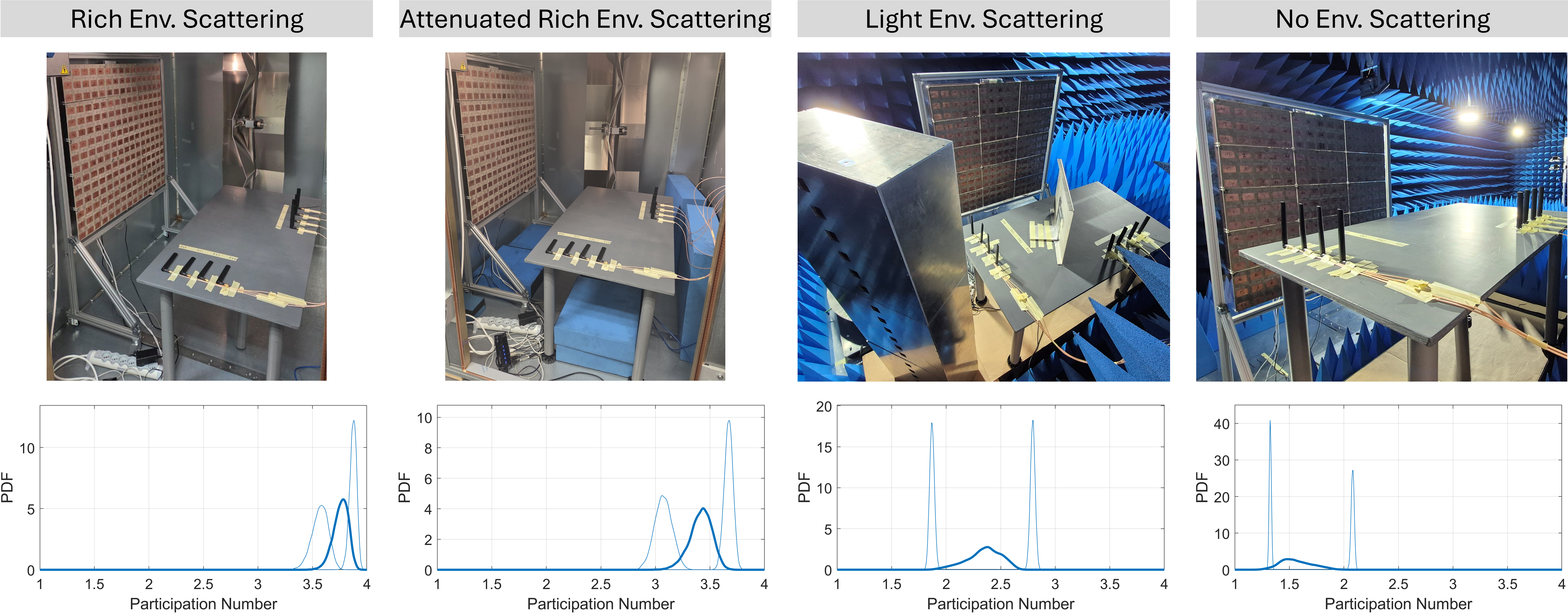}
    \caption{For each of the four considered radio environments, we display a photographic image in the top row and the PDFs of the number of BS-EEMDOFs for three distributions of $\mathbf{x}$ (\textit{RAND}, \textit{MAX}, \textit{MIN}; distinction by inspection) for the \textit{PIN} constraint on the entries of $\mathbf{r}$ in the bottom row. Each displayed PDF is evaluated based on 10,000 samples using a finite-difference evaluation of the Jacobian with respect to the RIS control variables, leveraging accurate predictions of the end-to-end MIMO channel matrix based on a proxy MNT model.}
    \label{Fig4}
\end{figure*}

The four radio environments in our experiments are displayed in the top row of Fig.~\ref{Fig4} and summarized as follows:
\begin{itemize}
    \item \textit{Rich Env. Scattering}: Our radio environment is a commercial reverberation chamber (RC, $1.75\ \mathrm{m}\times1.5\ \mathrm{m}\times2\ \mathrm{m}$).
    \item \textit{Attenuated Rich Env. Scattering}: Our radio environment is the same RC as before but loaded with absorbing material. 
    \item \textit{Light Env. Scattering}: Our radio environment is a commercial anechoic chamber (AC) containing a few metallic scattering objects. 
    \item \textit{No Env. Scattering}: Our radio environment is the same AC without any additional scattering objects. 
\end{itemize}
To be clear, the RC's mode stirrer seen in Fig.~\ref{Fig4} is not used, i.e., it is static throughout all experiments reported in this work. 

Our RIS prototype comprises 225 half-wavelength-sized RIS elements designed for operation around 2.45~GHz; we only use 100 RIS elements in this work (the remaining ones are always in a fixed reference configuration). The PIN diodes parametrizing the RIS elements are electrically very small, in line with the assumption of our system model (see Sec.~\ref{subsubsec_system_model}). Further details on the RIS design can be found in~\cite{ahmed2025over}.
Our transmit and receive arrays each consist of four parallel, half-wavelength-spaced antennas (ANT-W63WS2-SMA). In the RC setups, the orientations of the two arrays are perpendicular; in the AC setups, the orientations of the two arrays are parallel. 
Since our analysis is based on the end-to-end channel matrix, it can handle any choice of orientation of the antennas and any choice of radio environment.
We use an eight-port VNA (two cascaded Keysight P5024B 4-port VNAs) to measure $\mathbf{H}$ for a given RIS control vector at our operating frequency of 2.45~GHz.

Our results are displayed in Fig.~\ref{Fig4}. For each radio environment, the PDF of the number of BS-EEMDOFs for $\textit{RAND}$, $\textit{MAX}$ and $\textit{MIN}$ coherent illumination is shown (for the $\textit{PIN}$ constraint on the entries of $\mathbf{r}$); the corresponding means and standard deviations are summarized in Table~\ref{table2}. 
To begin, we observe that the average number of BS-EEMDOFs monotonically decreases as the scattering in the radio environment decreases. This observation is in line with  intuition derived from conventional EEMDOFs where rich scattering is known to boost the number of EEMDOFs~\cite{moustakas2000communication}.
Moreover, we observe that the distributions with optimized coherent illumination are markedly narrower in the AC-based radio environments. This makes sense because with fixed coherent illumination, the absence of MC between the backscatter elements collapses the distribution to a single value (as discussed in Sec.~\ref{subsubsec_examinationBSEEMDOFs}). 
Indeed, we expect the MC to be weak in the AC-based radio environments for our half-wavelength-sized RIS elements. As seen in~\cite{rabault2024tacit}, the environmental scattering has a strong influence on the MC strength between our RIS elements. 
Furthermore, we observe that the difference between \textit{MIN} and \textit{MAX} increases as the environmental scattering strength decreases. For rich environmental scattering, the average number of BS-EEMDOFs is 3.57 with \textit{MIN} and 3.87 with \textit{MAX}. The latter is close to the upper bound of $\tilde{N}=4$; the difference between \textit{MIN} and \textit{MAX} is only 0.3. Meanwhile, without any environmental scattering, the average number of BS-EEMDOFs is 1.33 with \textit{MIN} and 2.08 with \textit{MAX}. The former is close to the lower bound of unity; the difference between \textit{MIN} and \textit{MAX} is 0.75. Optimizing the coherent illumination of a backscatter array to maximize the number of BS-EEMDOFs appears thus particularly impactful in weakly scattering radio environments.

\begin{table}[t]
    \centering
    \caption{Mean and standard deviation of the distributions of the \\number of BS-EEMDOFs shown in Fig.~\ref{Fig4}.}
    \begin{tabular}{cccc}
        \hline
         & \textit{RAND} & \textit{MAX} & \textit{MIN} \\
        \hline
        \textit{Rich Env. Scattering} & $3.76 \pm 0.07$  & $3.87 \pm 0.03$ & $3.57 \pm 0.07$ \\
        \textit{Attenuated Rich Env. Scat.} & $3.42 \pm 0.10$ & $3.67 \pm 0.04$ & $3.07 \pm 0.08$  \\
        \textit{Light Env. Scattering} & $2.35 \pm 0.16$ & $2.79 \pm 0.02$ & $1.87 \pm 0.02$ \\
        \textit{No Env. Scattering} & $1.56 \pm 0.14$  & $2.08 \pm 0.01$ & $1.33 \pm 0.01$\\
        \hline
    \end{tabular}
    \label{table2}
\end{table}

\section{Conclusion}
\label{sec_conclusion}

To summarize, we have introduced a rigorous definition for the number of effective electromagnetic degrees of freedom in backscatter MIMO systems. In contrast to conventional MIMO systems, we found that the number of BS-EEMDOFs is generally a distributed variable (unless MC is negligible) that depends on the coherent illumination of the backscatter array. The illumination can thus be optimized to shape the distribution of the number of BS-EEMDOFs. Based on MNT, we derived a closed-form expression for the Jacobian underlying the BS-EEMDOF definition, revealing that the associated modes lie in the column space of the end-to-end channel matrix from the backscatter array ports to the receiver ports. 
We presented numerical and experimental results on the evaluation and optimization of the number of BS-EEMDOFs in various setups. The numerical full-wave simulation granted us direct access to all MNT parameters, allowing us to study the influence of various constraints on the loads and evaluating the benchmark number of EEMDOFs if the array is coherently fed rather than tunably terminated. The experimental results allowed us to systematically study the influence of the environmental scattering strength, while requiring care in handling the inevitable ambiguities in experimentally estimated proxy MNT parameters.

Looking forward, future research can explore optimizing other features of the distribution of BS-EEMDOFs (e.g., mean, variance) as well as the influence of the symbol duration (which is limited by the speed with which loads can be switched) and the quantization of available load states on the rate at which BS-MIMO systems can transfer information. Moreover, the presented analysis can be modified to study systems encoding input information into \textit{both} the input wavefront \textit{and} the load configuration~\cite{yan2019passive,basar2020reconfigurable,yan2020passive,lin2020reconfigurable,karasik2021adaptive}.
In addition, the presented analysis can be applied to rigorously examine the expressivity of wave-domain physical neural networks with structural non-linearity~\cite{momeni2023backpropagation}.

\appendix

In this Appendix, we describe the derivation of~(\ref{eq:JBS_full}). 

Applying the identity $\mathrm{d}\mathbf{A}^{-1} = -\mathbf{A}^{-1} \,(\mathrm{d}\mathbf{A})\, \mathbf{A}^{-1}$~\cite{hjorungnes2007complex} to $\mathbf{A}(\mathbf{r})=\mathbf{I}_{N_\mathrm{S}}-\mathbf{\Phi}(\mathbf{r})\,\mathbf{S}_\mathcal{SS}$, we obtain
\begin{equation}
 \mathrm{d}\mathbf{A}(\mathbf{r}) = -\mathrm{d}\mathbf{\Phi}(\mathbf{r})\, \mathbf{S}_\mathcal{SS} =  -\operatorname{diag}(\mathrm{d}\mathbf{r})\, \mathbf{S}_\mathcal{SS}. 
\end{equation}
Next, we apply the same identity to $\mathbf{G}(\mathbf{r})=\left(\mathbf{A}(\mathbf{r})\right)^{-1}$:
\begin{equation}
    \mathrm{d}\mathbf{G}(\mathbf{r})=-\mathbf{G}(\mathbf{r}) \, \mathrm{d}\mathbf{A}(\mathbf{r}) \, \mathbf{G}(\mathbf{r}) = \mathbf{G}(\mathbf{r}) \, \mathrm{diag}(\mathrm{d}\mathbf{r})\, \mathbf{S}_\mathcal{SS} \, \mathbf{G}(\mathbf{r}) .
\end{equation}
Then, we apply the product rule to $\mathbf{H}(\mathbf{r})=\mathbf{S}_\mathcal{RT} + \mathbf{S}_\mathcal{RS}\,\mathbf{G}(\mathbf{r})\,\mathbf{\Phi}(\mathbf{r})\,\mathbf{S}_\mathcal{ST}$, which yields 
\begin{equation}
\begin{aligned}
  \mathrm{d}\mathbf{H}(\mathbf{r}) &= \mathbf{S}_\mathcal{RS}
     \bigl( \mathrm{d}\mathbf{G}(\mathbf{r})\,\mathbf{\Phi}(\mathbf{r})\,\mathbf{S}_\mathcal{ST} + \mathbf{G}(\mathbf{r})\,\mathrm{d}\mathbf{\Phi}(\mathbf{r})\,\mathbf{S}_\mathcal{ST} \bigr)
     \\&= \mathbf{S}_\mathcal{RS}\,\mathbf{G}(\mathbf{r})\,\mathrm{diag}(\mathrm{d}\mathbf{r})
     \Bigl(        \mathbf{S}_\mathcal{SS}\,\mathbf{G}(\mathbf{r})\,\mathbf{\Phi}(\mathbf{r})\,\mathbf{S}_\mathcal{ST}       + \mathbf{S}_\mathcal{ST}      \Bigr) 
     \\&= \mathbf{S}_\mathcal{RS}\,\mathbf{G}(\mathbf{r})\,\mathrm{diag}(\mathrm{d}\mathbf{r})
     \,\mathbf{W}(\mathbf{r}), 
\end{aligned}
\end{equation}
where $\mathbf{W}(\mathbf{r})$ is defined as in (\ref{eq_def_W}).
Next, we apply the product rule to $\mathbf{y}(\mathbf{r},\mathbf{x})=\mathbf{H}(\mathbf{r})\mathbf{x}$, which yields
\begin{align}
  \mathrm{d}\mathbf{y}(\mathbf{r},\mathbf{x})
  &= \mathrm{d}\mathbf{H}(\mathbf{r})\,\mathbf{x} \nonumber\\
  &= \mathbf{S}_\mathcal{RS}\,\mathbf{G}(\mathbf{r})\,
     \mathrm{diag}(\mathrm{d}\mathbf{r})\,\mathbf{W}(\mathbf{r})\,\mathbf{x}.
\end{align}
Next, recognizing that $\mathbf{W}(\mathbf{r})\,\mathbf{x}\in\mathbb{C}^{N_\mathrm{S}}$ has the same size as $\mathbf{r}$, we apply the identity $\mathrm{diag}(\mathrm{d}\mathbf{a})\,\mathbf{b}      = \mathrm{diag}(\mathbf{b})\,\mathrm{d}\mathbf{a}$ that holds when the vectors $\mathbf{a}$ and $\mathbf{b}$ have the same size. We obtain
\begin{equation}
    \mathrm{d}\mathbf{y}(\mathbf{r},\mathbf{x}) = \mathbf{S}_\mathcal{RS}\,\mathbf{G}(\mathbf{r})\, \mathrm{diag}(\mathbf{W}(\mathbf{r})\,\mathbf{x})\,
     \mathrm{d}\mathbf{r}.
     \label{eq_dyyy}
\end{equation}
Finally, evaluating \eqref{eq_dyyy} at $\mathbf{r}=\mathbf{r}_0$ and comparing it to $\mathrm{d}\mathbf{y}(\mathbf{r}_0,\mathbf{x})  = \mathbf{J}(\mathbf{r}_0,\mathbf{x})\, \mathrm{d}\mathbf{r}$ directly yields~(\ref{eq:JBS_full}).

\section*{Acknowledgment}
The author acknowledges stimulating discussions with S.~S.~A.~Yuan. The author further acknowledges J.~Tapie as well as I.~Ahmed, F. Boutet, and C. Guitton, who, under the author's supervision, previously conducted the WNoC simulation for the work presented in~\cite{tapie2023systematic} and built the RIS prototype for the work presented in~\cite{ahmed2025over}, respectively. Moreover, the author acknowledges J.~Sol, who provided technical support for setting up the experiments.

\bibliographystyle{IEEEtran}

\providecommand{\noopsort}[1]{}\providecommand{\singleletter}[1]{#1}%

\end{document}